\documentclass[11pt]{iopart}

\usepackage{amsbsy,amssymb}

\usepackage{mathrsfs}

\newcommand{\al}{\alpha}
\newcommand{\be}{\beta}

\newcommand{\de}{\delta}
\newcommand{\m}{\mu}
\newcommand{\n}{\nu}

\newcommand{\si}{\sigma}
\newcommand{\cj}{\mathscr{J}}
\newcommand{\emrho}{\rho_e}

\newcommand{\Si}{\Sigma}

\newcommand{\mc}{\mathcal}
\newcommand{\ce}{\mc{E}}
\newcommand{\emce}{\mathscr{E}}

\newcommand{\cq}{\mc{Q}}
\newcommand{\emcb}{\mathscr{B}}
\newcommand{\ca}{\mc{A}}
\newcommand{\sch}{{\mbox{s}}}
\newcommand{\veh}{{\mbox{v}}}

\newcommand{\eps}{\epsilon}

\begin{document}

\title[1+1+2 Electromagnetic perturbations on non-vacuum LRS class II space-times]{1+1+2 Electromagnetic perturbations on non-vacuum LRS class II space-times: Decoupling scalar and 2-vector harmonic amplitudes}

\author{R. B. Burston}
\address{Max Planck Institute for Solar System Research,
37191 Katlenburg-Lindau, Germany}
\eads{\mailto{burston@mps.mpg.de}}

\begin{abstract}
We use the covariant and gauge-invariant 1+1+2 formalism of Clarkson and Barrett \cite{Clarkson2003} to analyze electromagnetic (EM)  perturbations on non-vacuum {\it locally rotationally symmetric} (LRS) class II space-times. Ultimately, we show how to derive six real decoupled equations governing the total of six  EM scalar and 2-vector harmonic amplitudes. Four of these are new, and result from expanding the complex EM 2-vector which we defined in \cite{Burston2007} in terms of EM 2-vector harmonic amplitudes. We are then able to show that there are four  precise combinations of the amplitudes that decouple, two of these are polar perturbations whereas the remaining two are axial.  The remaining two decoupled equations are the generalized Regge-Wheeler equations which were developed previously in \cite{Betschart2004}, and these govern the two EM scalar harmonic amplitudes. However, our analysis generalizes this by including a full description and classification of energy-momentum sources, such as charges and currents.
\end{abstract}

\maketitle

\section{Introduction}

There has been recent  interest \cite{Clarkson2003,Burston2007,Betschart2004,Clarkson2004} in the analysis of first-order perturbations to {\it locally rotational symmetric} (LRS) class II space-times \cite{Ellis1967,Stewart1968,Elst1996} using Clarkson and Barrett's 1+1+2 formalism \cite{Clarkson2003}.  Electromagnetic (EM) perturbations to non-vacuum LRS class II space-times via the 1+1+2 formalism were first analyzed in \cite{Betschart2004}.  Therein they derived covariant and gauge-invariant generalized Regge-Wheeler (RW) \cite{ Regge1957} equations governing the radial parts of the electric and magnetic fields, $\emce$ and $\emcb$. They also presented a new scalar and vector harmonic expansion formalism that naturally generalizes the {\it spherical} harmonic formalism developed in \cite{Clarkson2003}.  Finally, they presented a detailed analysis of the EM scalar harmonic amplitudes $\emce_\sch$ and $\emcb_\sch$, {\it in the absence of energy-momentum sources}. Following this, we showed that expressing Maxwell's equations in a 1+1+2 complex form is conducive to deriving a new decoupled, covariant and gauge-invariant, complex equation governing a new complex EM 2-vector \cite{Burston2007}, \mbox{$\Phi_\m := \emce_\m +\rmi \,\emcb_\m$}.  Thus clearly demonstrating, that the EM 2-vector fields decouple from the scalar fields.

In this paper, a vector harmonic expansion of the complex EM 2-vector is used and the governing equation yields a system of two equations which are coupled by the complex EM 2-vector harmonic amplitudes, $\Phi_\veh$ and $\bar \Phi_\veh$. We then discuss the invariance properties of this new coupled system and note that it is precisely analogous to the invariance properties of the 1+1+2 EM system we discussed in \cite{Burston2007}.  Therefore, precisely the same linear algebra techniques as in \cite{Burston2007} are used to choose new complex variables that are successful for decoupling the system.  It is then possible to separate the real from the imaginary parts and this ultimately results in four {\it real} decoupled equations governing four specific combinations of the {\it real} EM 2-vector harmonic amplitudes, $\emce_\veh$, $\emcb_\veh$, $\bar \emce_\veh$ and $\bar \emcb_\veh$, with a complete description of energy-momentum sources.

 Furthermore, this analysis is supplemented by the generalized RW equations as presented in \cite{Betschart2004}. However, the equations are generalized here by explicitly writing the scalar harmonic expansion of the generalized RW equations and also  include a treatment of the energy-momentum sources. Summarily, we find six {\it real} decoupled equations governing the six {\it real} scalar and 2-vector harmonic amplitudes, with a full scalar and 2-vector harmonic analysis of the energy-momentum source terms. Moreover, all quantities are neatly categorized into axial and polar perturbations. 
 
 Finally, we adhere to all notations and conventions as presented in \cite{Clarkson2003,Betschart2004} and this differs from that employed in \cite{Burston2007}.
\section{Previous Work}

\subsection{Betschart and Clarkson's 1+1+2 non-vacuum LRS class II space-time}

The 1+1+2 scalars, and equations, governing background LRS class II space-times were initiated in \cite{Clarkson2003} for the covariant Schwarzschild space-time, and later generalized by Betschart and Clarkson in \cite{Betschart2004} for arbitrary non-vacuum LRS class II space-times. There is a set of ten non-vanishing scalars given by, 
\begin{eqnarray}
\mbox{LRS class II:}\,\, \{\ca, \,\theta,\, \phi,\,\Si,\,\mc{E} ,\,\m,\, p,\,\cq,\,\Pi,\, \Lambda\}. \label{lrsIIscalr}
\end{eqnarray}
Here, $\ca $ is the radial acceleration of the four-velocity, $\theta$ and $\phi$ are respectively the expansions of the 3-sheets and 2-sheets, $\Si$ is the radial part of the shear of the 3-sheet, and $\ce$ is the radial component of the gravito-electric tensor. The energy-momentum quantities, mass-energy density, pressure, radial heat flux and radial anisotropic stress are denoted respectively $\m$, $p$, $\cq$ and $\Pi$, and finally, $\Lambda$ is the cosmological constant. These scalars do not vary over the 2-sheets and consequently, the covariant 2-derivative associated with the 2-sheets, which is denoted $\de_\m$, of any of these scalars will vanish.  Also, the equations governing these scalars arise from the Ricci identities and the Bianchi identities, and are reproduced here from  \cite{Betschart2004}: 

\begin{eqnarray}
\hat\phi+\frac 12\,\phi^2+\Bigl(\Si-\frac 23\,\theta\Bigr)\Bigl(\Si+\frac 13\,\theta\Bigr)+\ce=-\frac 23\,(\mu+\Lambda) -\frac 12\,\Pi, \label{rb1}\\
\hat\Si -\frac23\,\hat\theta+\frac 32\, \phi\,\Si=-\cq,\\
\hat \ce+\frac 32\,\phi\,\ce=\frac1 3\, \hat\m+\frac 12\,\Bigl(\Si -\frac 23 \,\theta\Bigr)\, \cq-\frac 12 \,\hat \Pi-\frac 34\, \phi\,\Pi,\\
\dot\phi+\Bigl(\Si-\frac 23\, \theta\Bigr)\Bigl(\ca-\frac12\,\phi\Bigr) =\cq,\\
\dot\Si -\frac23\,\dot\theta -\frac12 \,\Bigl(\Si-\frac 23 \,\theta\Bigr)^2 +\ca\,\phi+\ce= \frac13\,(\m+3\, p -2 \, \Lambda)+\frac12\,\Pi,\\
\fl\dot\ce -\frac 32\,\Bigl( \Si-\frac23\,\theta\Bigr)\ce =\frac 13\,\dot\m-\frac 12\, \dot\Pi +\frac 14\,\Bigl( \Si-\frac23\, \theta\Bigr)\Pi+\frac 12\, \phi\,\cq-\frac 12\,(\m+p)\Bigl(\Si-\frac 23\, \theta\Bigr),\\
\hat \ca+(\ca+\phi)\ca-\dot\theta-\frac 13\, \theta^2-\frac 32\, \Si^2= \frac 12 \,(\m+3\,p-2\,\Lambda),\\
\dot\m+\theta\,\m +\hat\cq+(2\,\ca+\phi)\cq+ \theta\,p+\frac 32\, \Si\, \Pi =0,\\
\dot\cq+\Bigl(\Si+\frac 43 \, \theta\Bigr)\cq+\hat p +\ca\,p +\hat \Pi+\Bigl(\ca+\frac 32\,\phi\Bigr)\Pi +\m \,\ca =0.\label{rblast}
\end{eqnarray}
Here, the ``dot" derivative is defined $\dot X_{\m\dots\n} := u^\al \nabla_\al X_{\m\dots\n}$ where $X_{\m\dots\n}$ represents any quantity. The ``hat" derivative is defined $\hat W_{\m\dots\n} := n^\al D_\al W_{\m\dots\n}$, where $W_{\m\dots\n}$ represents a 3-tensor and $D_\m$ is the covariant derivative associated with the 3-sheets. 

Furthermore, the Gaussian curvature of the 2-sheet is also reproduced from \cite{Betschart2004} is given by
\begin{eqnarray}
K=\frac 13\,(\m+\Lambda)-\ce-\frac 12 \,\Pi+\frac 14 \,\phi^2-\frac 14\,\Bigl(\Sigma-\frac 23\,\theta\Bigr)^2.\label{gauscurv}
\end{eqnarray}
Thus, the background system of evolution, propagation and transportation equations are given by \eref{rb1}-\eref{rblast} and the background scalars \eref{lrsIIscalr} are assumed to become known quantities in the forthcoming first-order perturbation equations.

\subsection{First-order Maxwell's equations in 1+1+2 complex form}

We now consider first-order EM perturbations on LRS class II space-times given by the EM fields ($E_\m$ and $B_\m$), charges ($\emrho$) and the current ($J_\m$). Furthermore, these perturbations are regarded as gauge-invariant according to the Sachs-Stewart-Walker Lemma \cite{Sachs1964,Stewart1974}. The first-order vectors are irreducibly split into 1+1+2 form according to
\begin{eqnarray}
\fl E_\m = \emce\, n_\m + \emce_\m ,\qquad B_\m = \emcb \, n_\m +\emcb_\m \qquad\mbox{and} \qquad J_\m = \cj n_\m +\cj_\m.
\end{eqnarray}
We showed in a recent paper \cite{Burston2007} that Maxwell's first-order equations may be expressed in a new 1+1+2 complex form which is conducive to decoupling,
\begin{eqnarray}
\hat \Phi +\phi\,\Phi+ \de^\al\Phi_\al=\emrho,\label{divons}\\
\dot\Phi  -\Bigl(\Si-\frac 23\, \theta\Bigr)\Phi+\rmi\,\eps^{\al\be}\,\de_\al\Phi_\be=- \cj,\label{divsec}\\
 \dot \Phi_{\bar \m}+\Bigl(\frac 12\Si+\frac 23 \theta\Bigr)\Phi_\m-\rmi{\epsilon_\m}^\al\Bigl[\hat \Phi_\al+\Bigl(\ca+\frac 12\phi\Bigr)\Phi_\al\Bigr]+\rmi{\eps_\m}^\al\de_\al \Phi=- \cj_\m,\label{thirdone}
\end{eqnarray}
where $\rmi$ is the complex number and $\eps_{\m\n}$ is the Levi-Civita 2-tensor. Moreover, the complex EM scalar and the complex EM 2-vector have been defined,
\begin{eqnarray}
\Phi     := \emce+\rmi \,\emcb \qquad\mbox{and} \qquad  \Phi_\m :=  \emce_\m+\rmi \,\emcb_\m. \label{phiphidef}
\end{eqnarray}
Subsequently, we showed in \cite{Burston2007} that the complex scalar and 2-vector naturally decouple. Furthermore, the gauge-invariant and covariant equations arise from the 1+1+2 complex system \eref{divons}-\eref{thirdone},
\begin{eqnarray}
 \fl \ddot \Phi   -\Bigl(\Si-\frac 53\, \theta\Bigr)\dot \Phi -\hat{\hat\Phi} -(\ca+2\,\phi)\hat \Phi -V\,\Phi=\mc{S},\label{GRW} \\
\fl \ddot \Phi_{\bar \m} - \Bigl(\Si-\frac 53 \,\theta\Bigr)\dot \Phi_{\bar\m}-\hat{\hat \Phi}_{\bar\m} -(\ca+2\,\phi)\, \hat \Phi_{\bar\m}-V_{{{(1)}}} \Phi_\m \nonumber\\
 \qquad -\rmi \,{\epsilon_\m}^\al \,\left[(2\,\ca-\phi)\dot \Phi_\al   -3 \, \Si\,\hat \Phi_\al -V_{(2)}\, \Phi_\al\right] = \mc{S}_\m,\label{waveforphia}
\end{eqnarray}
where terms related to the background potentials were defined
\begin{eqnarray}
\fl V :=\de^2+ 2\,K -\m+p+\Pi -2\,\Lambda ,\\
\fl V_{(1)}:=\de^2+\ce+\frac14\phi^2-\ca^2+\phi\ca +\frac 74\, \Si^2-\frac 29\,\theta^2+\frac 23\,\theta\,\Si-\frac 13\,\m+p-\frac 43\Lambda,\\
\fl V_{(2)} := -\dot\ca-\frac13\,\hat\theta-\frac 23\,\theta\,(\phi+2\,\ca) +\frac 12\,\Sigma\,(\phi+4\,\ca),
\end{eqnarray}
and the complex first-order energy-momentum sources are
\begin{eqnarray}
\fl \mc{S} :=  - \hat \emrho -(\phi+\ca) \emrho-\dot \cj  -\theta\,\cj +\rmi\,\epsilon^{\al\be} \de_\al \cj_\be,\\
\fl \mc{S}_\m:= -\dot \cj_{\bar\m}+\frac32\,\Bigl(\Si-\frac 23 \, \theta\Bigr) \cj_\m-\de_\m \emrho+\rmi\, {\epsilon_\m}^\al \left(\de_\al \cj -\hat \cj_\al-\frac32\,\phi\, \cj_\al \right), \label{phimsource}
\end{eqnarray}
where the 2-Laplacian is $\de^2 := \de^\al \de_\al$.  Thus, \eref{GRW}-\eref{waveforphia} clearly demonstrates the decoupling of the complex scalar and 2-vector. Moreover, \eref{GRW} is the generalized RW equation in complex form as presented in \cite{Burston2007}. Then a further decoupling between the EM scalars $\emce$ and$\emcb$ can be achieved by taking the real and imaginary parts  of \eref{GRW} separately, which then correspond to those previously derived in \cite{Betschart2004}. 

\subsection{Scalar and Vector Harmonics}

The scalar and vector harmonic expansion for all 1+1+2 first-order quantities has been previously presented in  \cite{Betschart2004} for non-vacuum LRS class II space-times and this was a natural generalization of the scalar and vector {\it spherical} harmonics given for the covariant Schwarzschild case in \cite{Clarkson2003}. Here we state the necessary results from \cite{Betschart2004}. Dimensionless sheet harmonic functions $Q$ (defined on the background) are reproduced here  from \cite{Betschart2004},
\begin{eqnarray}
\de^2 Q=-\frac{k^2}{r^2}\, Q, \qquad \hat Q= \dot Q =0,
\end{eqnarray}
where $k^2$ is real.  The scalar, $r$, is covariantly defined by the following differential equations
\begin{eqnarray}
\hat r -\frac12\,\phi \,r =0,\qquad \dot r  +\frac 12\,\Bigl(\Si-\frac23\,\theta\Bigr) \,r=0 \qquad  \mbox{and} \qquad \de _\m r =0.
\end{eqnarray}
Thus any first-order scalar function is expanded according to
\begin{eqnarray}
\psi = \sum_k \psi_\sch^{(k)} Q^{(k)} =\psi_\sch \, Q,
\end{eqnarray}
where the subscript, $\sch$,  indicates that a scalar harmonic expansion has been made. Furthermore, the summation over $k$ is implicit in the last equality and $\psi_\sch$ is referred to as the scalar harmonic amplitude, or  just scalar amplitude.

Similarly, all vectors are expanded in terms of even ($Q_\m$) and odd ($\bar Q_\m$) parity vector harmonics which are defined respectively
\begin{eqnarray}
Q_\m =r\, \de_\m Q  \qquad &&\rightarrow\qquad \de^2 Q_\m =\Bigl(K-\frac{k^2}{r^2}\Bigr) \, Q_\m , \label{evenqm}\\
\bar Q_\m =r\, {\eps_\m}^\al \de_\al Q  \qquad &&\rightarrow\qquad \de^2 \bar Q_\m =\Bigl(K-\frac{k^2}{r^2}\Bigr) \, \bar Q_\m .\label{oddqu}
\end{eqnarray}
However, there is a subtle difference with the equations presented here from \cite{Betschart2004}. Note that 2-Laplacian operating on the vector harmonics in \eref{evenqm}-\eref{oddqu} are left in terms of the Gaussian curvature, whereas in \cite{Betschart2004} they use  a further constraint of $K=\frac 1{r^2}$ which is equivalent to choosing a particular normalization that was convenient for their analysis. Here we leave this normalization general. 

Now, any first-order vector may be expanded according to
\begin{eqnarray}
\psi_\m =\sum_k \psi_\veh ^{(k)} \, Q_\m^{(k)} +\bar \psi_\veh ^{(k)} \, \bar Q_\m^{(k)} = \psi_\veh \, Q_\m +\bar \psi_\veh \, \bar Q_\m,
\end{eqnarray}
where the, $\veh$, is indicative of a 2-vector harmonic expansion. Also, the summation in the last quantity is implied, and $\psi_\veh$ and $\bar \psi_\veh$ are the 2-vector harmonic amplitudes or 2-vector amplitudes.

\section{Scalar Harmonic expansion of the complex RW equation}

A scalar (and vector) harmonic expansion of the EM scalars has been previously studied in \cite{Betschart2004} for vanishing energy-momentum sources. Here, our analysis generalizes those results by explicitly writing the generalized RW equation in a scalar harmonic form and furthermore, we include a treatment of energy-momentum sources.  The complex EM scalar, $\Phi$, and energy-momentum source, $\mc{S} $, are expanded in terms of scalar harmonics according to 
\begin{eqnarray}
\Phi = \Phi_\sch\, Q \qquad\mbox{and}\qquad  \mc{S}= \mc{S}_\sch\, Q.
\end{eqnarray}
Similarly, the first-order charges and currents become
\begin{eqnarray}
\emrho = {\emrho}_\sch\, Q, \qquad \cj = \cj_\sch \, Q \qquad \mbox{and}\qquad \cj_\m = \cj_\veh\,Q_\m + \bar \cj_\veh \, \bar Q_\m.
\end{eqnarray}
Therefore, the complex generalized RW equation \eref{GRW} becomes
\begin{eqnarray}
 \ddot \Phi_\sch   -\Bigl(\Si-\frac 53\, \theta\Bigr)\dot \Phi_\sch -\hat{\hat\Phi}_\sch -(\ca+2\,\phi)\hat \Phi_\sch -V\,\Phi_\sch=\mc{S}_\sch.\label{GRWsch}
\end{eqnarray}
where the  potential and source are now
\begin{eqnarray}
 V =-\frac{k^2}{r^2}+ 2\,K -\m+p+\Pi -2\,\Lambda ,\\
 \mc{S}_\sch=  - \hat {\emrho}_\sch -(\phi+\ca) {\emrho}_\sch-\dot \cj_\sch  -\theta\cj_\sch +\rmi\, \frac{k^2}{r} \bar \cj_\veh.
\end{eqnarray}
Therefore, since $\Phi_\sch$ and $\mc{S}_\sch$ are the {\it only} complex quantities in \eref{GRWsch}, the real and imaginary parts can be taken separately to obtain two real decoupled equations,
\begin{eqnarray}
\fl  \ddot \emce_\sch   -\Bigl(\Si-\frac 53\, \theta\Bigr)\dot \emce_\sch -\hat{\hat\emce}_\sch -(\ca+2\,\phi)\hat \emce_\sch -V\,\emce_\sch=- \hat {\emrho}_\sch -(\phi+\ca) {\emrho}_\sch-\dot \cj_\sch  -\theta\cj_\sch, \label{schce} \\
 \fl  \ddot \emcb_\sch   -\Bigl(\Si-\frac 53\, \theta\Bigr)\dot \emcb_\sch -\hat{\hat\emcb}_\sch -(\ca+2\,\phi)\hat \emcb_\sch -V\,\emcb_\sch=\frac{k^2}{r} \bar \cj_\veh .\label{sxchebb}
\end{eqnarray}
Furthermore, we can see that the perturbations neatly separate into polar and axial perturbations which  was similarly noted in \cite{Betschart2004} and are respectively
\begin{eqnarray}
\mbox{Decoupled polar perturbation:}\qquad\{\emce_\sch\},\\
\mbox{Decoupled axial perturbation:} \qquad\{\emcb_\sch \}.
\end{eqnarray}
Here, we also further categorize the energy-momentum sources  according to
\begin{eqnarray}
\mbox{Energy-momentum polar perturbations:}\qquad\{ {\emrho}_\sch, \cj_\sch\}, \label{emcat}\\
\mbox{Energy-momentum axial perturbations:}\qquad\{\bar \cj_\veh \}.\label{emcat1}
\end{eqnarray}
At this stage it is also clear that the remaining energy-momentum quantity ($\cj_\veh$) must be a polar perturbation since its partner ($\bar \cj_\veh$) is axial. We will see that this is consistent with the vector harmonic expansions in the following section.

As a final note, we inspect the energy-momentum source terms in \eref{schce} and \eref{sxchebb} and note that the polar electric scalar amplitude, $\emce_\sch$, is being forced purely by the polar energy-momentum scalar amplitudes, ${\emrho}_\sch$ and  $\cj_\sch$. Furthermore, the  axial magnetic scalar amplitude, $\emcb_\sch$, is being forced purely by the axial current 2-vector amplitude, $\bar \cj_\veh$, and completely vanishes in the cases where the harmonic summation index, $k^2$, also vanishes.

\section{Vector Harmonic expansion}

Much more work is required to decouple the EM 2-vector amplitudes, $\emce_\veh$ and $\emcb_\veh$. First, the complex EM 2-tensor \eref{phiphidef} and the energy-momentum source \eref{phimsource} are expanded according to
\begin{eqnarray}
\Phi_\m=\Phi_\veh \, Q_\m+ \bar \Phi_\veh \,\bar Q_\m  \qquad\mbox{and}\qquad \mc{S}_\m = \mc{S}_\veh \, Q_\m + \bar {\mc{S}} \,\veh \,\bar Q_\m \label{Phiexpsh}.
\end{eqnarray}
Thus, by substituting \eref{Phiexpsh} into \eref{waveforphia} we get a coupled system of the form
\begin{eqnarray}
\ddot \Phi_\veh- \Bigl(\Si-\frac 53 \,\theta\Bigr)\dot \Phi_\veh-\hat{\hat \Phi}_\veh -(\ca+2\,\phi)\, \hat \Phi_\veh-V_{{{(1)}}} \Phi_\veh \nonumber\\
 \qquad +\rmi \,\left[(2\,\ca-\phi)\dot {\bar \Phi}_\veh   -3 \, \Si\,\hat {\bar \Phi}_\veh  -V_{(2)}\,\bar  \Phi_\veh\right]=\mc{S}_\veh,\label{Phivone}\\
 \ddot {\bar \Phi}_\veh- \Bigl(\Si-\frac 53 \,\theta\Bigr)\dot{ \bar \Phi}_\veh-\hat{\hat {\bar \Phi}}_\veh -(\ca+2\,\phi)\, \hat{ \bar\Phi}_\veh-V_{{{(1)}}} \bar \Phi_\veh \nonumber\\
 \qquad -\rmi \,\left[(2\,\ca-\phi)\dot \Phi_\veh   -3 \, \Si\,\hat \Phi_\veh -V_{(2)}\, \Phi_\veh\right]=\bar \mc{S}_\veh.\label{Phivtwo}
\end{eqnarray}
The background potential and energy momentum sources similarly become
\begin{eqnarray}
\fl V_{(1)}:=-\frac{k^2}{r^2}+K+\ce+\frac14\phi^2-\ca^2+\phi\ca +\frac 74\, \Si^2-\frac 29\,\theta^2+\frac 23\,\theta\,\Si-\frac 13\,\m+p-\frac 43\Lambda,\\
\fl  \mc{S}_\veh:= -\dot \cj_\veh+\frac32\,\Bigl(\Si-\frac 23 \, \theta\Bigr) \cj_\veh-\frac 1r \, {\emrho}_\sch+\rmi\, \left(  \hat{\bar  \cj}_\veh+\frac32\,\phi\, {\bar \cj}_\veh \right) , \\
\fl \bar \mc{S}_\veh:=-\dot {\bar \cj}_\veh+\frac32\,\Bigl(\Si-\frac 23 \, \theta\Bigr) \bar \cj_\veh+\rmi\,\left(\frac1r\,\cj_\sch -\hat \cj_\veh-\frac32\,\phi\, { \cj}_\veh \right) .
\end{eqnarray}
By inspection, equations \eref{Phivone}-\eref{Phivtwo} are (momentarily ignoring sources) now precisely invariant under the simultaneous transformation of the form $\Phi_\veh \rightarrow \bar \Phi_\veh$  and $\bar\Phi_\veh\rightarrow- \Phi _\veh$. Thus, this is precisely the same invariance exhibited by the 1+1+2 EM system as discussed in \cite{Burston2007} and decoupling may be achieved by choosing new independent variables  and sources according to 
\begin{eqnarray}
\Phi_- := \Phi_\veh- \rmi\, \bar \Phi_\veh\qquad\mbox{and}\qquad \Phi_+ := \Phi_\veh+ \rmi\, \bar \Phi_\veh, \label{phipphim}
\end{eqnarray}
and similarly for the source, $\mc{S}_{\pm} := \mc{S}_\veh \pm \rmi\, \bar {\mc{S}}_\veh$, where the ``$\pm$" is relative. Consequently, by taking complex additions and subtractions of  \eref{Phivone}-\eref{Phivtwo}  we find two decoupled complex equations; one for each of $\Phi_+$ and $\Phi_-$, 
\begin{eqnarray}
\fl \ddot \Phi_+- \left[\Si-\frac 53 \,\theta - (2\,\ca-\phi)\right]\dot \Phi_+-\hat{\hat \Phi}_+ -(\ca+2\,\phi+3\,\Si)\, \hat \Phi_+-[V_{(1)} + V_{(2)}] \Phi_+=\mc{S}_+, \label{2baby1}\\
\fl \ddot \Phi_-- \left[\Si-\frac 53 \,\theta + (2\,\ca-\phi)\right]\dot \Phi_--\hat{\hat \Phi}_- -(\ca+2\,\phi-3\,\Si)\, \hat \Phi_--[V_{(1)} - V_{(2)}] \Phi_-=\mc{S}_-. \label{2baby2}
\end{eqnarray}
Therefore, since the only complex quantities in \eref{2baby1}-\eref{2baby2} are $\Phi_+$, $\Phi_-$ $\mc{S}_+$ and $\mc{S}_-$, the real and imaginary parts may be considered separately and there are actually four {\it real} decoupled equations; one for each of the real and imaginary parts of $\Phi_\pm$, i.e. $\Re[\Phi_+]$, $\Im[\Phi_+]$, $\Re[\Phi_-]$ and $\Im[\Phi_-]$.  These quantities can now be related back to the harmonic amplitudes of the real EM 2-vector. By using \eref{phipphim}, and noting that according to \eref{phiphidef}  
\begin{eqnarray}
\Phi_\veh := \emce_\veh + \rmi \, \emcb \qquad\mbox{and}\qquad \bar\Phi_\veh := \bar \emce_\veh + \rmi \, \bar \emcb,
\end{eqnarray}
then the real and imaginary parts of $\Phi_\pm$ are
\begin{eqnarray}
\fl \Phi_+ = (\emce_\veh - \bar \emcb_\veh) + \rmi \,( \emcb_\veh + \bar \emce_\veh)\qquad\mbox{and}\qquad \Phi_- = (\emcb_\veh+\bar \emce_\veh) + \rmi \,( \emcb_\veh - \bar \emce_\veh) , \label{pohdsifemd}
\end{eqnarray}
and the sources 
\begin{eqnarray}
\fl  \mc{S}_+:=-\frac 1r \, ({\emrho}_\sch+\cj_\sch) -\dot \cj_\veh+\hat \cj_\veh+\frac32\,\Bigl(\Si-\frac 23 \, \theta+\phi\Bigr) \cj_\veh \nonumber\\
-\rmi\, \left[ \dot {\bar \cj}_\veh- \hat{\bar  \cj}_\veh-\frac32\,\Bigl(\Si-\frac 23 \, \theta +\phi\Bigr) \bar \cj_\veh \right] , \label{spm1} \\
\fl  \mc{S}_-:=-\frac 1r \, ({\emrho}_\sch-\cj_\sch) -\dot \cj_\veh-\hat \cj_\veh+\frac32\,\Bigl(\Si-\frac 23 \, \theta-\phi\Bigr) \cj_\veh \nonumber\\
+\rmi\, \left[ \dot {\bar \cj}_\veh+ \hat{\bar  \cj}_\veh-\frac32\,\Bigl(\Si-\frac 23 \, \theta -\phi\Bigr) \bar \cj_\veh \right] . \label{spm2}
\end{eqnarray}
Therefore by inspecting  \eref{pohdsifemd}, it is now clear that we have four different combinations of the EM vector amplitudes which decouple. Furthermore, they may also be categorized into polar and axial perturbations. Summarily, the precise combinations of the harmonic amplitudes of the real EM 2-vectors that individually decouple are:
\begin{eqnarray}
\mbox{Decoupled polar perturbations:}\qquad\{\emce_\veh + \bar \emcb_\veh, \emce_\veh - \bar \emcb_\veh\}, \label{dafcsdVF}\\
\mbox{Decoupled axial perturbations:} \qquad\{\emcb_\veh + \bar \emce_\veh, \emcb_\veh - \bar \emce_\veh\}.\label{dafcsdVF1}
\end{eqnarray}
Furthermore, this also elucidates that the remaining energy-momentum quantity yet to be categorized is a polar perturbation as anticipated
\begin{eqnarray}
\mbox{Energy-momentum polar perturbation:} \qquad\{ \cj_\veh\}.
\end{eqnarray}
Moreover, by inspecting \eref{dafcsdVF}-\eref{dafcsdVF1}, it is clear that if one were to integrate the four decoupled equations \eref{2baby1}-\eref{2baby2}, it is then possible to construct linear combinations of the solutions to ultimately find each of $\emce_\veh$, $\emcb_\veh$, $\bar \emce_\veh$ and $\bar \emcb_\veh$.

\section{Summary}

We have provided a complete analysis of the EM perturbations to the most general non-vacuum LRS class II space-times, using the complex 1+1+2 formalism and a harmonic scalar and vector expansion of the complex EM scalar and 2-vector.  We derived four new decoupled equations which govern four precise combinations of the EM 2-vector amplitudes and this has a full energy-momentum treatment. Furthermore, we generalized the two RW equations, which govern the scalar amplitudes, from \cite{Betschart2004} to also include energy-momentum quantities. This gives a total of six decoupled equations governing the combined total of  six EM amplitudes. Summarily, the necessary combinations  that individually decouple are:
\begin{eqnarray}
\mbox{Decoupled polar perturbations:}\qquad\{\emce_\sch, \emce_\veh + \bar \emcb_\veh, \emce_\veh - \bar \emcb_\veh\}, \\
\mbox{Decoupled axial perturbations:} \qquad\{\emcb_\sch,\emcb_\veh + \bar \emce_\veh, \emcb_\veh - \bar \emce_\veh\}.
\end{eqnarray}
Furthermore, the energy-momentum sources also fall into polar and axial perturbations according to
\begin{eqnarray}
\mbox{Energy-momentum polar perturbations:}\qquad\{ {\emrho}_\sch, \cj_\sch, \cj_\veh\}, \\
\mbox{Energy-momentum axial perturbations:}\qquad\{\bar \cj_\veh \}.
\end{eqnarray}

We present the main results, i.e. the decoupled equations \eref{schce}-\eref{sxchebb} and \eref{2baby1}-\eref{2baby2},  as a single master equation of the form 
\begin{eqnarray}
\fl \ddot \Psi- \left[\Si-\frac 53 \,\theta -  \si\,(2\,\ca-\phi)\right]\dot \Psi-\hat{\hat \Psi} -(\ca+2\,\phi +\si\,3 \,\Si)\, \hat \Psi-U\, \Psi=\mathscr {S},
\end{eqnarray}
where for 
\subsubsection*{polar perturbations:} 
\begin{eqnarray}
\fl \Psi = \emce_\sch&\qquad \mbox{for}\qquad  \si =0,\qquad U =V  \qquad& \mbox{and}\qquad\mathscr {S}= \Re[\mc{S}_\sch] ,\\
\fl  \Psi = (\emce_\veh - \bar \emcb_\veh)   &\qquad \mbox{for}\qquad  \si =1,\qquad U = V_{(1)} + V_{(2)}\qquad &\mbox{and}\qquad \mathscr {S}= \Re[\mc{S}_+],\\
\fl  \Psi = (\emce_\veh + \bar \emcb_\veh)   &\qquad \mbox{for}\qquad  \si =1,\qquad U = V_{(1)} + V_{(2)}\qquad &\mbox{and}\qquad\mathscr {S}=\Re[\mc{S}_-].
\end{eqnarray}
\subsubsection*{axial perturbations:}
\begin{eqnarray}
\fl \Psi = \emcb_\sch &\qquad \mbox{for}\qquad  \si =0, \qquad \,\,\,\,U =V \qquad & \mbox{and}\qquad\mathscr {S}= \Im[\mc{S}_\sch],\\
\fl  \Psi = ( \bar \emcb_\veh+\emce_\veh ) &\qquad \mbox{for}\qquad  \si =-1,\qquad U = V_{(1)} - V_{(2)}\qquad&\mbox{and}\qquad\mathscr {S}= \Im[\mc{S}_+] ,\\
\fl  \Psi = ( \emcb_\veh -  \bar \emce_\veh)  &\qquad \mbox{for}\qquad  \si =-1,\qquad U= V_{(1)} -V_{(2)}\qquad&\mbox{and}\qquad \mathscr {S}=\Im[\mc{S}_-] .
\end{eqnarray}

It is anticipated that the decoupling analysis provided here will be useful for modeling space-times for which there are small EM fields, charges and currents present. Moreover, we also note that this analysis is the primary motivation to decouple the significantly harder problem of {\it gravitational} perturbations on LRS class II space-times using the {\it gravitoelectromagnetic} (GEM) formalism \cite{Bel1958}. It is very well established that the GEM equations governing gravitational fields are remarkably similar to the EM equations governing EM fields \cite{Maartens1998}. In a later paper, we will show that the EM 2-vector harmonic amplitudes behave in a very similar manner to the GEM 2-tensor amplitudes, and consequently, we decouple four precise combinations of the GEM 2-tensor harmonic amplitudes for gravitational perturbations.

\end{document}